%% file: paper.tex
\documentclass[prl,twocolumn,showpacs,showkeys,preprintnumbers,floatfix,nofootinbib,superscriptaddress]{revtex4-1}
\usepackage{amsfonts} 
\usepackage{amssymb} 
\usepackage{amsmath} 
\usepackage{graphicx} 
\usepackage{subfigure} 
\usepackage{array} 
\usepackage{dcolumn} 
\usepackage{bm} 
\usepackage{latexsym} 
\usepackage{longtable} 
\usepackage{hyperref} 
\usepackage{bbold}
\usepackage{color}
\usepackage{multirow}
\usepackage{rotating}
\usepackage{comment}
\usepackage{bm}
\usepackage[utf8]{inputenc} 

\providecommand{\abs}[1]{\lvert#1\rvert}
\providecommand{\matrixe}[3]{\langle#1\lvert#2\rvert#3\rangle}
\providecommand{\rAsq}{\langle r_A^2 \rangle}
\providecommand{\CL}{\nonumber\\}

\graphicspath{{./figs/}}
\allowdisplaybreaks

\include{macro}

\begin{document}


\title{Axial Vector Form Factors from Lattice QCD that Satisfy the PCAC Relation}

\author{Yong-Chull Jang}
\email{ypj@bnl.gov}
\affiliation{Brookhaven National Laboratory, Physics Department, Upton, New York 11973}

\author{Rajan Gupta}
\email{rajan@lanl.gov}
\affiliation{Los Alamos National Laboratory, Theoretical Division T-2, Los Alamos, NM 87545}

\author{Boram Yoon}
\email{boram@lanl.gov}
\affiliation{Los Alamos National Laboratory, Computer Computational and Statistical Sciences, CCS-7, Los Alamos, NM 87545}

\author{Tanmoy Bhattacharya}
\email{tanmoy@lanl.gov}
\affiliation{Los Alamos National Laboratory, Theoretical Division T-2, Los Alamos, NM 87545}

\preprint{LA-UR-19-29302}
%
%
%
\date{\today}
\begin{abstract}
Previous lattice QCD calculations of axial vector and pseudoscalar
form factors show significant deviation from the partially conserved
axial current (PCAC) relation between them.  Since the original
correlation functions satisfy PCAC, the observed deviations from the
operator identity cast doubt on whether all the systematics in the
extraction of form factors from the correlation functions are under
control. We identify the problematic systematic as a missed excited
state, whose energy as a function of the momentum transfer squared,
$Q^2$, is determined from the analysis of the 3-point functions
themselves. Its mass is much smaller than those of the excited states
previously considered and including it impacts the extraction of all
the ground state matrix elements. The form factors extracted using
these mass/energy gaps satisfy PCAC and other consistency conditions,
and validate the pion-pole dominance hypothesis. We also show that the
extraction of the axial charge $g_A$ is very sensitive to the value of
the mass gaps of the excited states used and current lattice data do
not provide an unambigous determination of these, unlike the $Q^2 \neq
0$ case.  To highlight the differences and improvement between the
conventional versus the new analysis strategy, we present a comparison
of results obtained on a physical pion mass ensemble at $a\approx
0.0871$~fm. With the new strategy, we find $g_A = 1.30(6)$.  A very
significant improvement over previous lattice results is found for the
axial charge radius $r_A = 0.74(6) $~fm, extracted using the
$z$-expansion to parameterize the $Q^2$ behavior of $G_A(Q^2)$, and
$g_P^\ast = 8.06(44)$ obtained using the pion pole-dominance ansatz to
fit the $Q^2$ behavior of the induced pseudoscalar form factor
$\widetilde{G}_P(Q^2)$.
\end{abstract}
\maketitle


The nucleon axial form factor $G_A(Q^2)$ is an important input needed
to calculate the cross-section of neutrinos off nuclear targets. It is
not well-determined experimentally~\cite{Bernard:2001rs}, and the most
direct measurements using liquid hydrogen targets are unlikely to be
performed due to safety concerns. At present, these form factors are
typically extracted from measurements of scattering off nuclear
targets and involves modeling of nuclear
effects~\cite{Carlson:2014vla,AguilarArevalo:2010zc}, which introduces
uncertainties~\cite{Hill:2017wgb}.  Lattice QCD provides the best
approach to calculate these from first principles, however, one has to
demonstrate that all systematics are under control.

The axial, $G_A$, and the induced pseudoscalar, $\widetilde{G}_P$, form factors  are extracted
from the 
matrix elements of the four
components of the isovector axial current $A_\mu \equiv \overline{u} \gamma_5   \gamma_\mu d$ 
between the ground state of the nucleon: 
\begin{multline}
\label{eq:AFFdef}
\left\langle N({\bm p}_f) | A_\mu (\vec{q}) | N({\bm p}_i)\right\rangle  = \\
{\overline u}_N({\bm p}_f)\left( G_A(q^2) \gamma_\mu
+ q_\mu \frac{\tilde{G}_P(q^2)}{2 M}\right) \gamma_5 u_N({\bm p}_i),
\end{multline}
and the pseudoscalar form factor $G_P$ from 
\begin{equation}
\left\langle N({\bm p}_f) | P (\vec{q}) | N({\bm p}_i)\right\rangle  = 
{\overline u}_N({\bm p}_f) G_P(q^2) \gamma_5 u_N({\bm p}_i) \,, 
\label{eq:PSdef}
\end{equation}
where $P = \overline{u} \gamma_5 d$ is the pseudoscalar density,
$N({\bm p})$ is the nucleon state with mass $M$ and lattice momentum
${\bm p} \equiv 2 \pi {\bm n} / La$ with ${\bm n} \equiv (n_1, n_2, n_3)$.  We neglect the induced tensor
form factor $\tilde{G}_T$ in Eq.~\eqref{eq:AFFdef} since we assume
isospin symmetry, $m_u=m_d$,
throughout~\cite{Bhattacharya:2011qm}. All the form factors
will be presented as functions of the spacelike four-momentum transfer
$Q^2 \equiv {\bf p}^2 - (E-M)^2 = -q^2$.

In our previous work~\cite{Rajan:2017lxk}, we showed
that form factors with good statistical precision can be obtained from lattice
simulations, however, these data do not satisfy the partially conserved axial current (PCAC) relation:
\begin{equation}
2 {\widehat m} G_P(Q^2) = 2 M G_A(Q^2) - \frac{Q^2}{2M} {\tilde G}_P(Q^2) \,,
\label{eq:PCAC}
\end{equation}
where ${\widehat m}$ is the PCAC quark mass.  Such a failure has also
been observed in all other lattice 
calculations~\cite{Bali:2014nma,Green:2017keo,Alexandrou:2017hac,Capitani:2017qpc,Ishikawa:2018rew,Shintani:2018ozy}. 
Since PCAC is
an operator relation, it is important to find the systematic
responsible for the deviation, and remove it prior to comparing lattice data 
with phenomenology.\looseness-1

In this work we show that the problematic systematic is a missed lower energy
excited state. Using data from a
physical pion mass ensemble, $a09m130W$~\cite{Jang:2019jkn}, we show
how the mass and energy gap of this state can be
determined from the analysis of nucleon 3-point correlation
functions. We then demonstrate that form factors extracted including
these parameters satisfy PCAC and other consistency conditions.
With these improvements, we claim that the combined uncertainty in the
lattice data is reduced to below ten percent level.

All lattice data presented here are from our calculations using the
clover-on-HISQ formulation~\cite{Rajan:2017lxk,Jang:2019jkn}. The
gauge configurations are from the physical-mass $2+1+1$-flavor HISQ
ensemble $a09m130W$ generated by the MILC
collaboration~\cite{Bazavov:2012xda} with lattice spacing $a\approx
0.0871\,\fm$ and $M_\pi = 130\,\MeV$. The pion mass on these configurations with the clover valence quark action
is $M_\pi\approx 138\,\MeV$.  Further
details of the lattice parameters and methodology, statistics, the
interpolating operator used to construct the 2- and 3-point
correlation functions can be found in
Refs.~\cite{Jang:2019jkn,Rajan:2017lxk}.

\begin{figure*}[!htb]
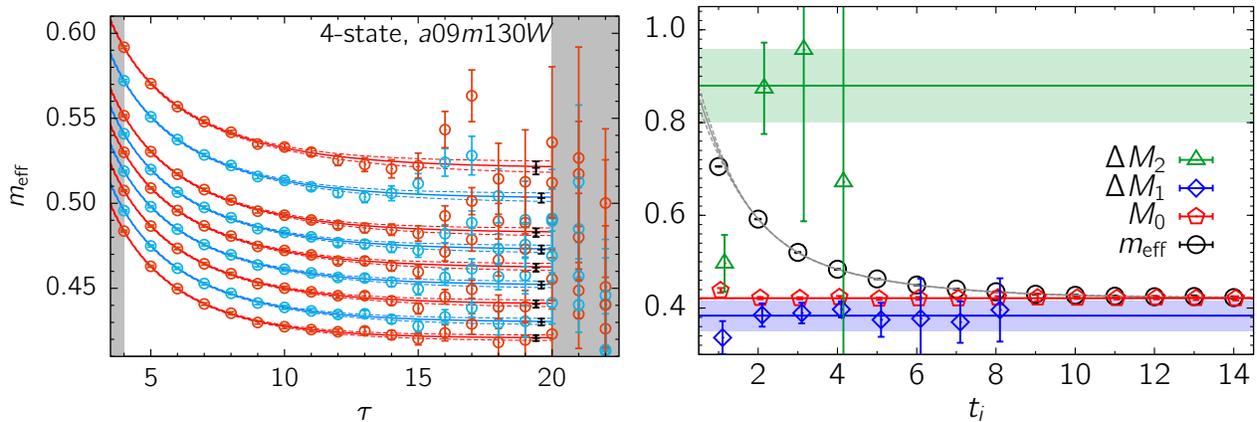
  
  \centering
  \includegraphics[width=0.94\columnwidth]{{{figs/meff_allmom}}}\hspace{2mm}
  \includegraphics[width=0.94\columnwidth]{{{2pt-prony/E.s3.prior1}}}
  \caption{The left panel shows the effective energy $E_{\rm
      eff}(\tau) = \log \frac{C^{\rm 2pt}(\tau)}{C^{\rm 2pt}(\tau+1)}$
    and the 4-state fit for various momentum channels.  The right
    panel compares the nucleon $M_{\rm eff}$ at $\bm p= 0$ with $M_0$
    (red band), the first mass gap $\Delta M_1 \equiv M_1 - M_0$ (blue
    band), and the second mass gap $\Delta M_2 \equiv M_2 - M_0$
    (green band). These are obtained using the Prony's method with
    fits to the intervals $[t_i, t_i+6]$, where $t_i$ is the starting
    time slice.  Sensitivity to $\Delta M_1$ ($\Delta M_2 $) is lost
    at $t_i = 8$ ($t_i = 4$). All data are in lattice units. }
  \label{fig:2pt}
\end{figure*}

The nucleon operator used to create and annihilate the nucleon state 
couples to the ground and all the excited and multiparticle states with 
appropriate quantum numbers. To isolate the ground state matrix elements, 
we fit the data for the 2- and 3-point functions, $C^\text{2pt}$ and $C_\Gamma^\text{3pt}$,  using 
their spectral decompositions. For the 2-point functions, the four states truncation is
\begin{align}
C^\text{2pt}(\tau,\bm{p}) =& 
  {|{\cal A}_0|}^2 e^{-E_0 \tau} + {|{\cal A}_1|}^2 e^{-E_1 \tau}\,+ \nonumber \\
  &{|{\cal A}_2|}^2 e^{-E_2 \tau} + {|{\cal A}_3|}^2 e^{-E_3 \tau}\,, 
\label{eq:2pt}
\end{align}
where ${\cal A}_i$ are the amplitudes and $E_i$ are the energies with
momentum $\bm{p}$.  The data and fits using Eq.~\eqref{eq:2pt} are
shown in Fig.~\ref{fig:2pt} (left). There is a reasonable plateau at
large $\tau$ in $M_{\rm eff}(\tau) \equiv \log \frac{C^{\rm
    2pt}(\tau)}{C^{\rm 2pt}(\tau+1)}$ for all momenta up to ${\bm n}^2
= 6$. The right panel shows $M_{\rm eff}$, $M_0$, and the first two
mass gaps, determined using a variant of the Prony's
method~\cite{Fleming:2009wb}, that are consistent with those obtained
from 4-state fits~\cite{Jang:2019jkn}.

The two-state truncation of the 3-point functions $C_\Gamma^{(3\text{pt})}
(t;\tau;\bm{p}^\prime,\bm{p})$, with Dirac index $\Gamma$, is 
\begin{align}
  C_\Gamma^{3\text{pt}}&(t;\tau;\bm{p}^\prime,\bm{p}) = \CL
   &\abs{\mathcal{A}_0^\prime} \abs{\mathcal{A}_0}\matrixe{0^\prime}{\mathcal{O}_\Gamma}{0} e^{-E_0t - M_0(\tau-t)} + \CL
   &\abs{\mathcal{A}_0^\prime} \abs{\mathcal{A}_1}\matrixe{0^\prime}{\mathcal{O}_\Gamma}{1} e^{-E_0t - M_1(\tau-t)} + \CL
   &\abs{\mathcal{A}_1^\prime} \abs{\mathcal{A}_0}\matrixe{1^\prime}{\mathcal{O}_\Gamma}{0} e^{-E_1t -M_0(\tau-t)} + \CL
   &\abs{\mathcal{A}_1^\prime} \abs{\mathcal{A}_1}\matrixe{1^\prime}{\mathcal{O}_\Gamma}{1} e^{-E_1t - M_1(\tau-t)}  \,,
   \label{eq:3pt}
\end{align}
where the source point is translated to $t=0$, the operator is
inserted at time $t$, and the nucleon is annihilated at the sink time
slice $\tau$. In this relation, $|0\rangle$ and $|n\rangle$ are the
ground and $n^{\rm th}$ excited state. The superscript ${}^\prime$
denotes that the state could have nonzero momentum $\bm{p}^\prime$. The
momentum transfer $\bm q = \bm p^\prime - \bm p = \bm p^\prime$ since
$\bm{p}$ at the sink is fixed to zero.  The $M_i$, $E_i$ and ${\cal
  A}_i^\prime|A_i$ are the masses, energies and the amplitudes for the
creation$|$annihilation of these states by the nucleon interpolating operator.

To display and discuss the data, it is much more convenient and common
to consider the five ratios, ${\cal R}_{5\mu}$ and ${\cal R}_{5}$, of
the 3-point correlation functions of $A_\mu$ and $P$ to the 2-point correlator as 
defined in Ref.~\cite{Rajan:2017lxk}:
\begin{align}
  {\cal R}_{51} \rightarrow&\; \frac{1}{\sqrt{(2 E_p (E_p+M))}} \left[ -\frac{q_1 q_3}{2M} {\tilde G}_P \right] \,,
  \label{eq:r2ff-GPGA1} \\
  {\cal R}_{52} \rightarrow&\; \frac{1}{\sqrt{(2 E_p (E_p+M))}} \left[ -\frac{q_2 q_3}{2M} {\tilde G}_P \right] \,,
  \label{eq:r2ff-GPGA2} \\
  {\cal R}_{53} \rightarrow&\; \frac{1}{\sqrt{(2 E_p (E_p+M))}} \left[ -\frac{q_3^2}{2M} {\tilde G}_P + (M+E_p) G_A \right]  \,,
  \label{eq:r2ff-GPGA3} \\
  {\cal R}_{54} \rightarrow&\; \frac{ 4M q_3}{\sqrt{(2 E_p (E_p+M))}} \left[ \frac{M-E_p}{2M} G_P + G_A \right] \,,
  \label{eq:r2ff-GPGA4} \\
  {\cal R}_{5} \rightarrow&\; \frac{1}{\sqrt{(2 E_p (E_p+M))}} \left[ q_3 {G}_P \right] \,. 
  \label{eq:r2ff-GP} 
\end{align}
where $q_i$ is the momentum transferred by the current in the ``i''
spatial direction. The direction ``3'' is singled out since it is
chosen to be the direction of the spin projection of the Dirac spinors
in the construction of the 3-point functions.  In the limit of large
source-sink separation, $\tau \to \infty$, these ratios give the
combination of the form factors shown on the right hand side. We have
explicitly displayed the kinematic factors to show which momentum
combinations have a signal in each case. Data with equivalent momenta
are averaged in the analysis.

\begin{figure*}[!htb]  
  \centering
  \includegraphics[width=0.94\columnwidth]{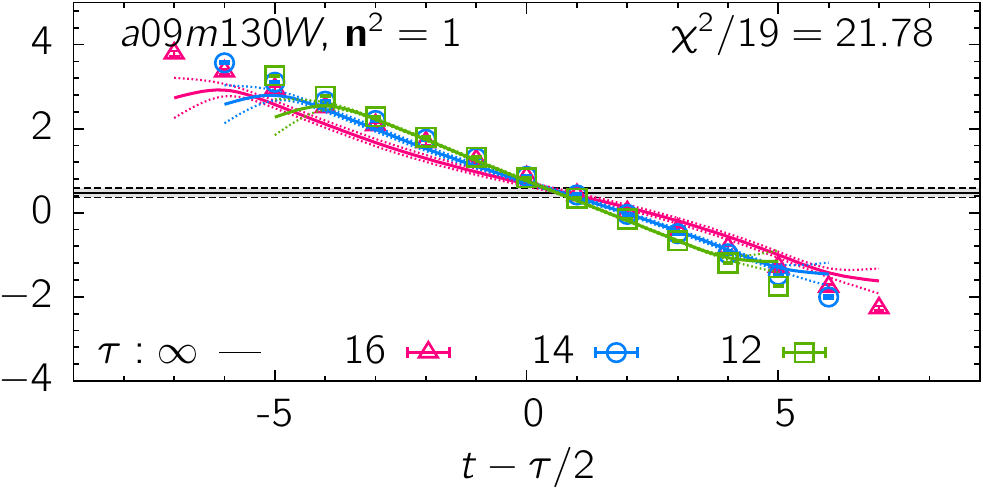}\hspace{2mm}
  \includegraphics[width=0.94\columnwidth]{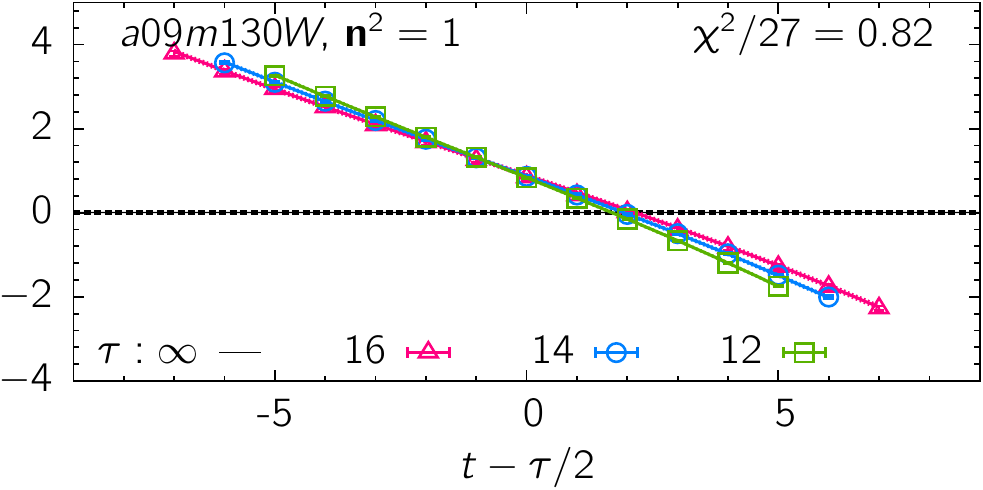}
  \caption{Comparison of the fits used to remove ESC in the $A_4$ 3-point
    function using the ${\cal S}_{\rm 2pt}$ (left) and ${\cal S}_{\rm
      A4}$ (right) strategies defined in the text. This data for
    ${\bm p} = (1,0,0)2\pi/La$ show the largest ESC. The values of $\tau$ and $\chi^2$/DOF 
    are given in the legend.}
  \label{fig:ESC}
\end{figure*}

Equations~\eqref{eq:r2ff-GPGA1}--\eqref{eq:r2ff-GPGA4} form an
over-complete set.  ${\cal R}_{51}$ and ${\cal R}_{52}$ can be
averaged as they are related by the lattice cubic symmetry and give
$\widetilde{G}_P$.  For $q_3=0$, ${\cal R}_{53}$ gives $G_A$. For the
other momentum combinations, one gets a linear combination of $G_A$
and $\widetilde{G}_P$. Thus, the three $A_i$ correlators give results for $G_A$  and ${\tilde G}_P$ 
for all values of momentum transfer. Consequently, data from $A_4$ correlators have
traditionally~\cite{Bali:2014nma,Rajan:2017lxk,Green:2017keo,Alexandrou:2017hac,Capitani:2017qpc,Ishikawa:2018rew,Shintani:2018ozy}
been neglected because they exhibits very large excited-state
contamination (ESC) as shown in Fig.~\ref{fig:ESC}.  Lastly, $G_P$ is
obtained uniquely from Eq.~\eqref{eq:r2ff-GP}.

In our previous
work~\cite{Rajan:2017lxk}, the energies of the excited state used to
isolate the ground state matrix elements in fits to the 3-point
functions were taken from four-state fits to the 2-point correlation
function defined in Eq.~\eqref{eq:2pt}. The resulting form factors
violated PCAC. Furthermore, the violation increased as $Q^2 \to 0$, $a
\to 0$ and $M_\pi \to M_\pi^{\rm physical}$. Correcting for $O(a)$ lattice artifacts
in the axial current showed no significant reduction in the violation~\cite{Rajan:2017lxk}.

In this paper we show that by using these values of $M_i$ and $E_i$ to
remove the ESC we missed a lower excited state. Furthermore, the
energy of this state can be determined from the analysis of the $A_4$
correlator, {\it ie}, the channel with the largest ESC is the most sensitive
to it. 

\begin{figure}[!tb]  
  \includegraphics[width=0.94\columnwidth]{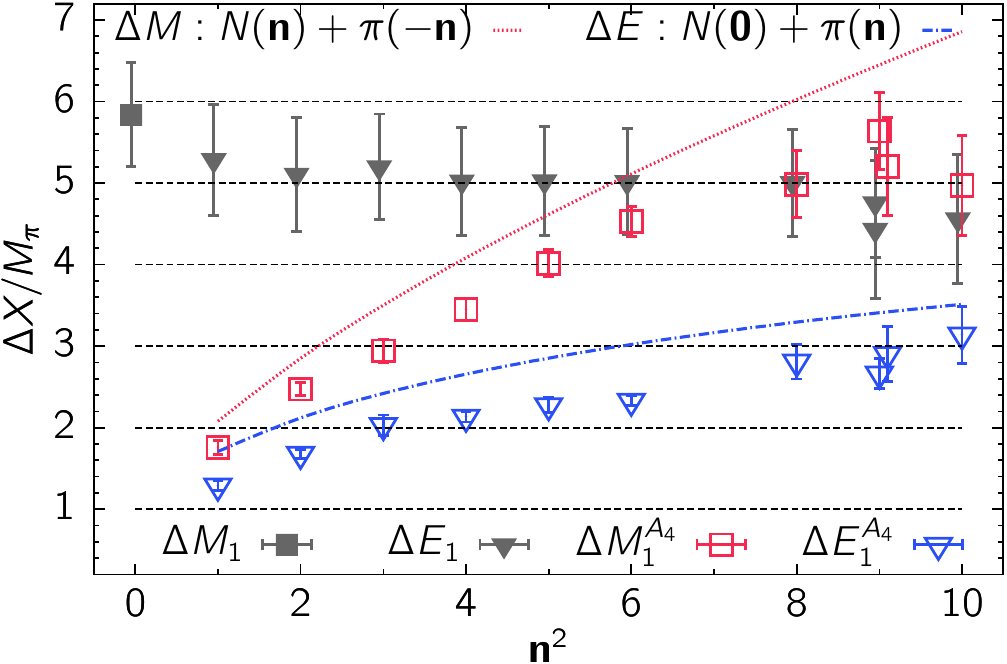}
  \caption{Mass and energy gaps $\Delta M_1 = M_1-M_0$ and $\Delta
    E_1(\bm{n}^2) = E_1^\text{2pt}-E_0$ in units of $M_\pi$ are obtained from the 4-state
    fit to the 2-point correlator. The values $\Delta M_1^{A_4}$ and
    $\Delta E_1^{A_4}$ for the $a09m130W$ ensemble are obtained using
    strategy ${\cal S}_{\rm A4}$. The dotted lines show the
    corresponding values for non-interacting $N({\bm
      p}) \pi({\bm -p})$ and $N(0) \pi({\bm p})$ states. }
  \label{fig:massgaps}
\end{figure}

\begin{figure}[!htb] 
  \centering
  \includegraphics[width=0.91\columnwidth]{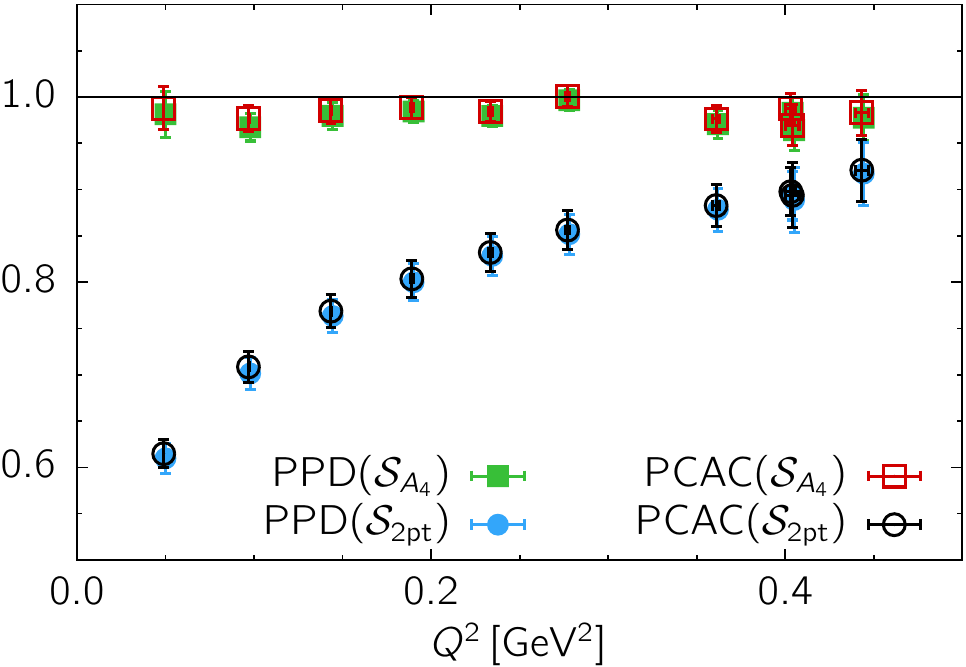}
  \caption{Comparison of the tests of the PCAC and the pion-pole
    dominance (PPD) hypothesis using the two analysis strategies
    ${\cal S}_{\rm A4}$ and ${\cal S}_{\rm 2pt}$ defined in the
    text. We plot the quantity $\frac{2 {\widehat m}}{2M}
    \frac{G_P(Q^2)}{G_A(Q^2)} + \frac{Q^2}{4M^2} \frac{{\tilde
        G}_P(Q^2)}{G_A(Q^2)}$ for PCAC (open symbols) and $
    \frac{Q^2+M_\pi^2}{4M^2} \frac{{\tilde G}_P(Q^2)}{G_A(Q^2)} $
    (filled symbols) for PPD. These should both be unity up to $O(a)$
    corrections at all $Q^2$ if these relations hold. }
  \label{fig:PCAC}
\end{figure}

In Fig.~\ref{fig:ESC}, we compare the conventional 3${}^\ast$-state
fit to the $A_4$ correlator with masses and energies, $M_i$ and $E_i$,
taken from the 4-state fit to the 2-point
function~\cite{Rajan:2017lxk} versus the new two-state fit with $M_1$
and $E_1$ left as free parameters. The $\chi^2/{\rm DOF}$ and
$p$-value of the fits for all ten momentum cases are given in
Table~\ref{tab:CHIsq}. Note that for ${\bm n}=(0,0,1)$, $\chi^2/{\rm
  DOF}$ reduces from 21.8 to 0.8. The values of the mass$|$energy gaps
of the ``first'' excited state shown in Fig.~\ref{fig:massgaps} are
much smaller, and close to those expected for non-interacting $N({\bm
  p}) \pi(-{\bm p})$ and $N(0) \pi({\bm p})$ states~\cite{Bar:2018xyi}.  By ${\bm n}^2
\gtrsim 6$, the mass gaps become similar and, correspondingly, the
violation of PCAC at larger momentum-transfer are observed to be small
(see Fig.~\ref{fig:PCAC} and Ref.~\cite{Rajan:2017lxk}). We
hypothesize that this lower energy excited state provides the dominant
contamination in all five 3-point correlation functions. On the basis
of consistency checks including PCAC, we make the case that it is
essential to include this lower energy excited state in all the fits
used to remove the ESC. \looseness-1

\begin{table}[!tbh]
  \caption{The goodness of the fits to $C^{\rm 3pt}_{A_4}$. The new 2-state fits 
   correspond to strategy ${\cal S}_{\rm A4}$ defined in the text. 
   The conventional 3$^\ast$-state fit values (strategy ${\cal S}_{\rm 2pt}$) 
   are taken from Refs.~\protect\cite{Jang:2018djx,Gupta:2018qil}.}
  \label{tab:CHIsq}
  \begin{ruledtabular}
    \begin{tabular}{l|cc|cc}
      & \multicolumn{2}{c|}{New 2-state} & \multicolumn{2}{c}{Conventional 3$^\ast$-state}  \\
      $\bm{n}^2$ & $\chi^2/\text{d.o.f}$  & $p$ & $\chi^2/\text{d.o.f}$  & $p$  \\
      \hline
      1                  & 0.817  & 0.73  &  21.78  & $< 5\times 10^{-5}$ \\
      2                  & 1.314  & 0.13  &  19.36  & $< 5\times 10^{-5}$ \\
      3                  & 1.263  & 0.16  &  11.79  & $< 5\times 10^{-5}$ \\
      4                  & 0.778  & 0.79  &  4.757  & $< 5\times 10^{-5}$ \\
      5                  & 1.268  & 0.16  &  5.348  & $< 5\times 10^{-5}$ \\
      6                  & 1.712  & 0.01  &  4.834  & $< 5\times 10^{-5}$ \\
      8                  & 0.815  & 0.74  &  1.724  & 0.03 \\
      9$\phantom{\prime}$ (2,2,1)          & 1.865  & 0.01  &  2.726  & 0.001 \\
      9$^\prime$ (3,0,0) & 0.539  & 0.98  &  0.974  & 0.49 \\
      10                 & 0.865  & 0.67  &  1.089  & 0.35 
    \end{tabular}
  \end{ruledtabular}
\end{table}

To highlight the differences and improvements, we define two
analysis strategies, ``conventional'', ${\cal S}_{\rm 2pt}$, and ``new'',
${\cal S}_{\rm A4}$: 
\begin{itemize}
\item
${\cal S}_{\rm 2pt}$: All the ground and excited state 
  $M_i$ and $E_i$, are taken from 4-state fits to the nucleon 2-point
  function and used in the $3^\ast$-state analysis of all the 3-point
  functions as detailed in Ref.~\cite{Jang:2019jkn}.
\item
${\cal S}_{\rm A4}$: The ground state parameters $M_0$ and $E_0$ are
  taken from the 4-state fits to the nucleon 2-point function. These are
  considered reliable based on the observed plateau in the effective-mass data at large $\tau$ 
as shown in Fig.~\ref{fig:2pt}.  The parameters for the first excited
  state, $M_1$ and $E_1$, are taken from fits to the $A_4$ 3-point
  correlator as discussed above. These are then used in a two-state
  analysis of all other 3-point functions.
\end{itemize}
In both cases, it is important to note that residual ESC may still be
present in the $M_i$ and $E_i$. Future higher precision
calculations will improve the precision of the calculations 
by steadily including more states in the fits. 

In Fig.~\ref{fig:massgaps}, we show three sets of data for the energy
gaps of the first excited state: $\Delta E_1^{\rm 2pt} \equiv E_1^{\rm
  2pt} - E_0$ obtained from fits to the 2-point correlator. These are
compared with the two values on either side of the $A_4$ operator
insertion, which are expected to be different since the correlator is
projected to zero-momentum at the sink: $\Delta M_1^{A_4} \equiv M_1^{A_4} - M_0$, 
the zero momentum case on the sink side and the
non-zero-momentum values  $\Delta E_1^{A_4} \equiv E_1^{A_4} - E_0$ on
the source side.  It is clear that $\Delta E_1^{A_4}$ and $\Delta
M_1^{A_4} $ are much smaller than $\Delta E_1^{\rm 2pt}$ for ${\bm
  n}^2 \lesssim 6$ indicating the contribution of a lower energy excited
state. Secondly, $\Delta E_1^{A_4}$ and $\Delta M_1^{A_4} $ are
significantly different. Strategy ${\cal S}_{\rm 2pt}$ corresponds to
using $\Delta E_1^{\rm 2pt}$ and $\Delta M_1^{\rm 2pt}$, whereas ${\cal S}_{\rm A4}$ corresponds
to using $\Delta E_1^{A_4}$ and $\Delta M_1^{A_4} $. \looseness-1

In Fig.~\ref{fig:massgaps}, we also show, using dotted lines, the
expected values for $\Delta E$ and $\Delta M$ if we assume that the
leading contribution of the current $A_4(\bm q)$ is to insert or
remove a pion with momentum $\bm q$. Thus the plotted $\Delta E$
corresponds to the values for a non-interacting state
$N({\bm p}=0) \pi({\bm p})$, while $\Delta M_1$ to $N({\bm p})
\pi(-{\bm p})$. In calculating these values, we have used the lattice
values for $M_0$ and $M_\pi$ and the relativistic dispersion relation, 
which is consistent with the data from the 2-point function.
The values and variation of $\Delta E_1^{A_4}$ and $\Delta M_1^{A_4} $
with ${\bm n}^2$ are roughly consistent with this picture.

\begin{figure*}[!htb] 
  \centering
  \subfigure{
  \includegraphics[width=0.95\columnwidth]{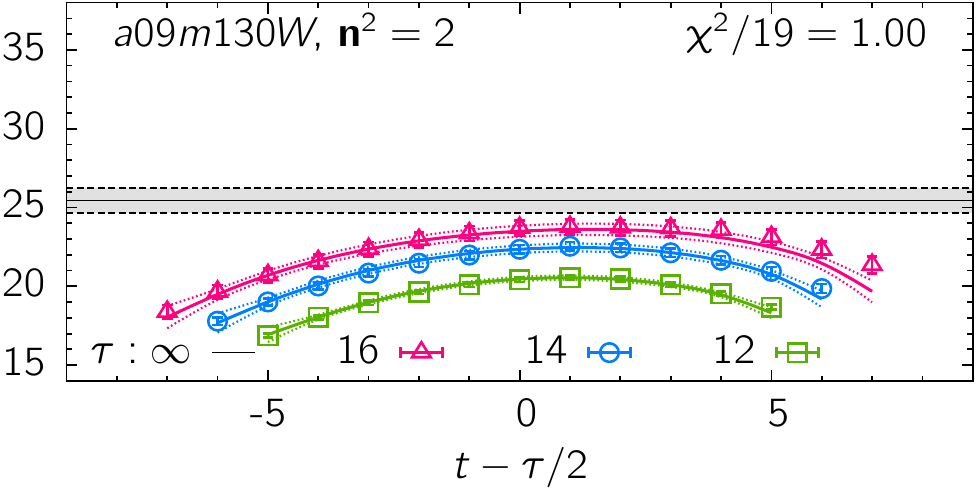}\hspace{6mm}
  \includegraphics[width=0.95\columnwidth]{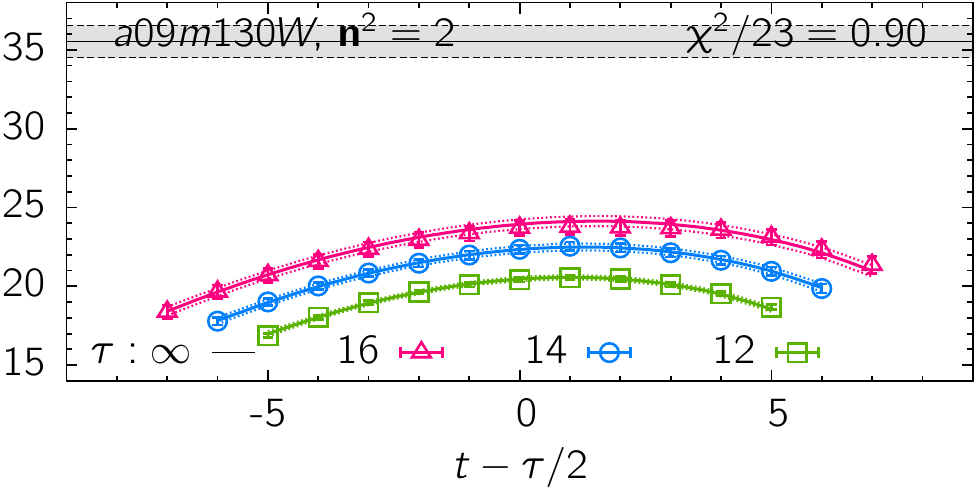}
  }
  \subfigure{
  \includegraphics[width=0.95\columnwidth]{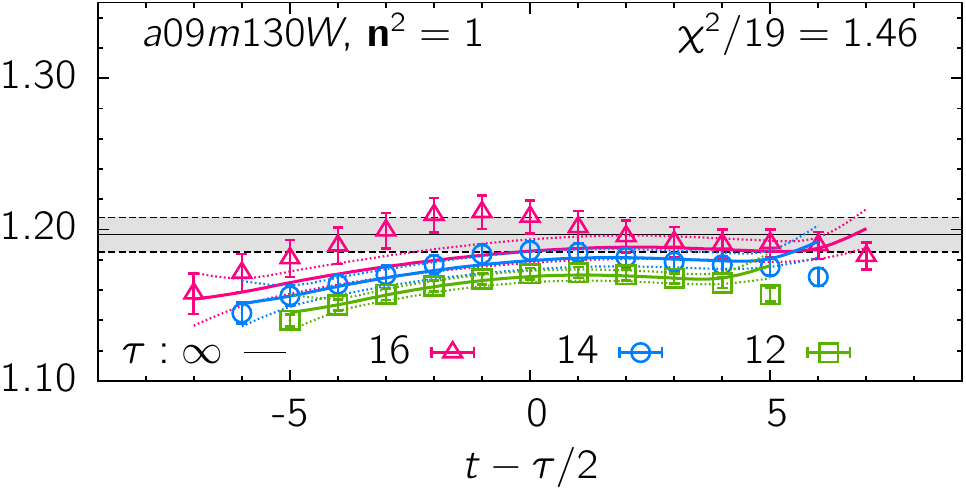}\hspace{6mm}
  \includegraphics[width=0.95\columnwidth]{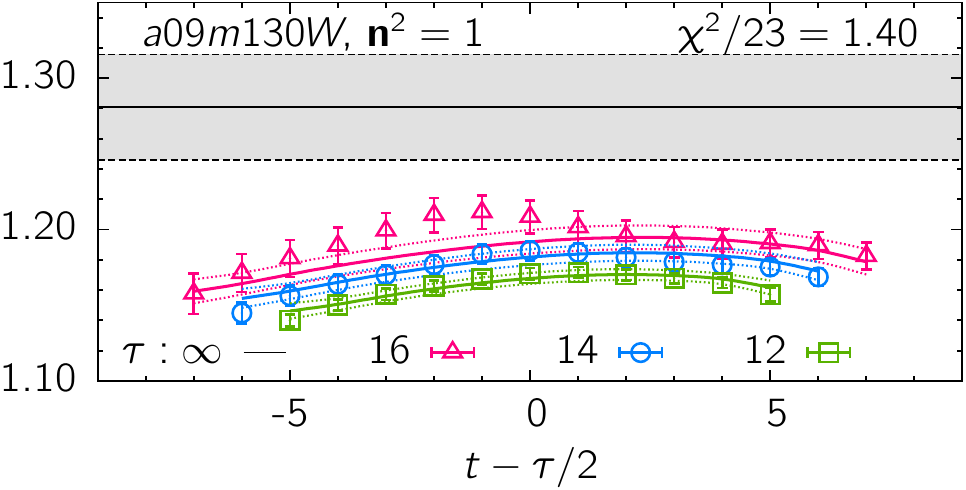}
  }
  \subfigure{
  \includegraphics[width=0.95\columnwidth]{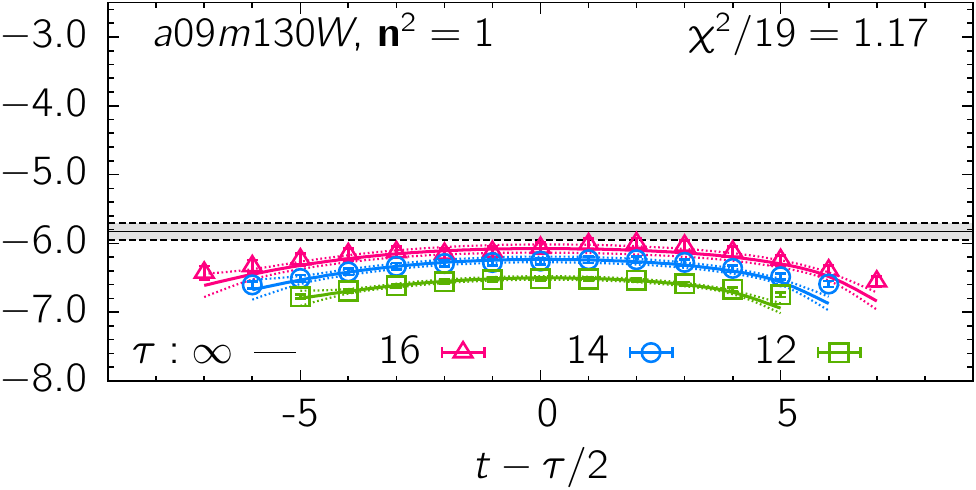}\hspace{6mm}
  \includegraphics[width=0.95\columnwidth]{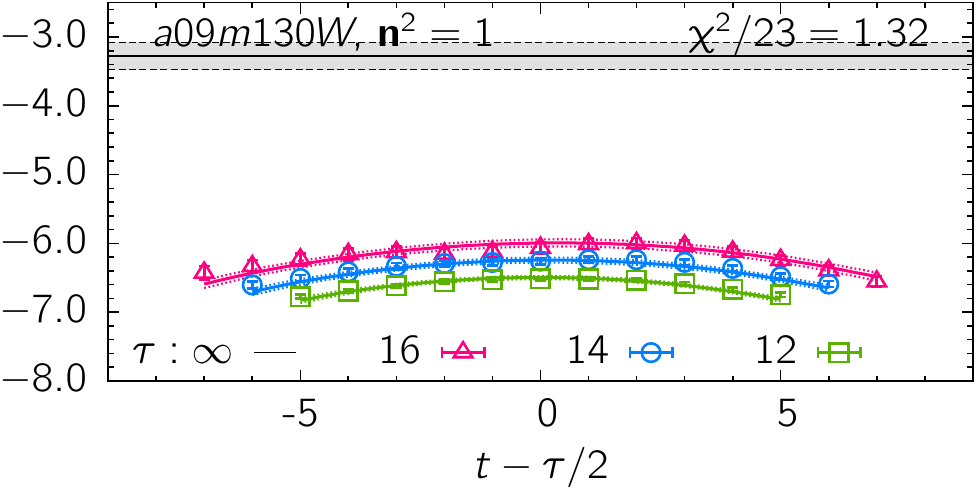}
  }
  \subfigure{
  \includegraphics[width=0.95\columnwidth]{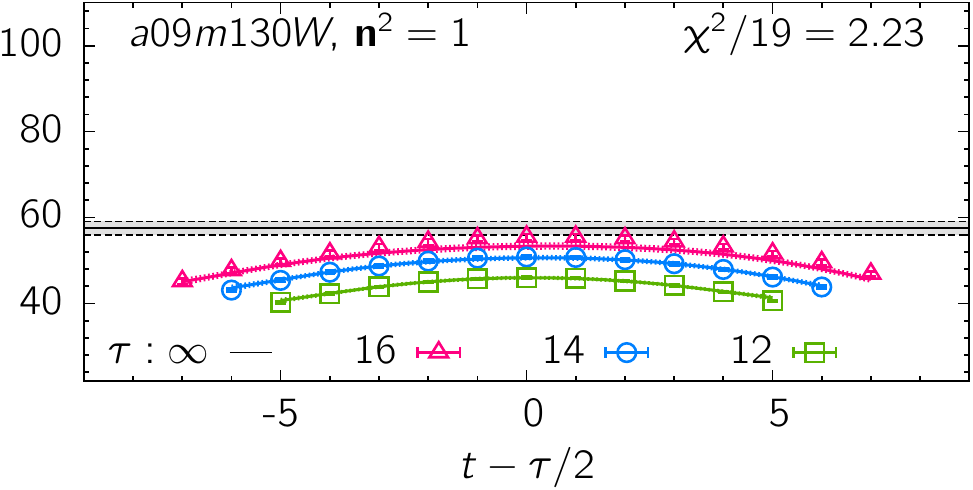}\hspace{6mm}
  \includegraphics[width=0.95\columnwidth]{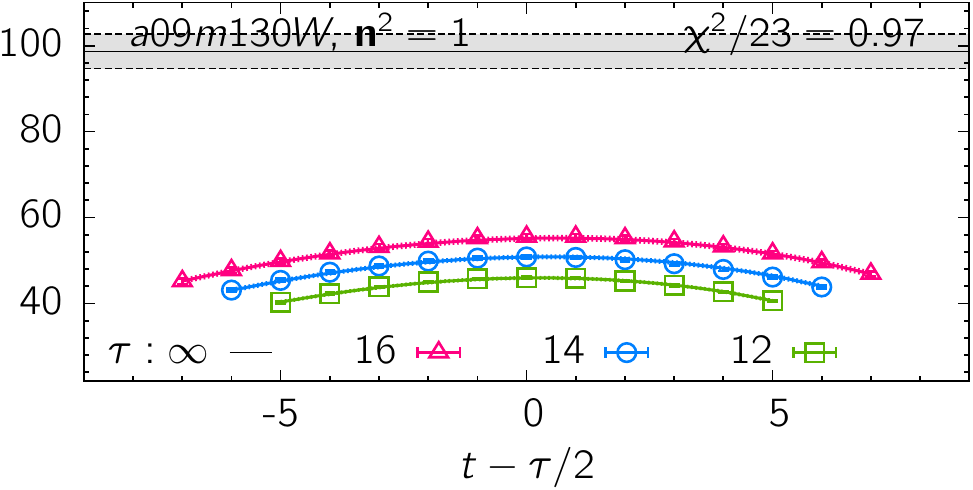}
  }
  \caption{Comparison of the ground state matrix element (the
    horizontal grey band) obtained using the two strategies ${\cal
      S}_{\rm 2pt}$ (left panels) and ${\cal S}_{\rm A4}$ (right
    panels). In both cases, a 2-state fit is performed to the 3-point
    correlator.  The four sets of figures are for: $A_1$ with ${\bm n}
    = (1,1,0)$ (top); $A_3$ with $q_3 = 0$ and ${\bm n} = (1,0,0)$
    (second row); $A_3$ with ${\bm n} = (0,0,1)$ (third row); and $P$
    with ${\bm n} = (1,0,0)$ (bottom row). The $\chi^2/$DOF and the
    values of $\tau$ used in the fit are shown in the legend.  }
  \label{fig:ESCcomp}
\end{figure*}

Using the excited-state parameters extracted from the analysis of the
$A_4$ correlator, and following the strategy ${\cal S}_{\rm A4}$ gives
very different values for the ground state matrix elements extracted
from the three spatial, $A_i$, and the $P$ correlators. A comparison
of the 2-state fits using ${\cal S}_{\rm A4}$ and the $3^\ast$-state
fit using ${\cal S}_{\rm 2pt}$ is shown in Fig.~\ref{fig:ESCcomp} 
for the lowest non-zero momentum channels. Based on the $\chi^2/$DOF
of the fits, we cannot distinguish between the two strategies except for the $P$ channel in spite of 
having high statistics data (165K measurements on 1290
configurations)~\cite{Jang:2019jkn}. The key point in each of the four channels is 
the convergence--it is from below and including the ``new'' lower excited
state (${\cal S}_{\rm A4}$) gives significantly larger values of the
matrix elements and thus the form factors. This pattern persists for ${\bm n}^2 \lesssim 5$, 
above which the difference in the mass gap does not have a significant effect.

\begin{figure*}[!htb] 
  \centering
  \includegraphics[width=0.63\columnwidth]{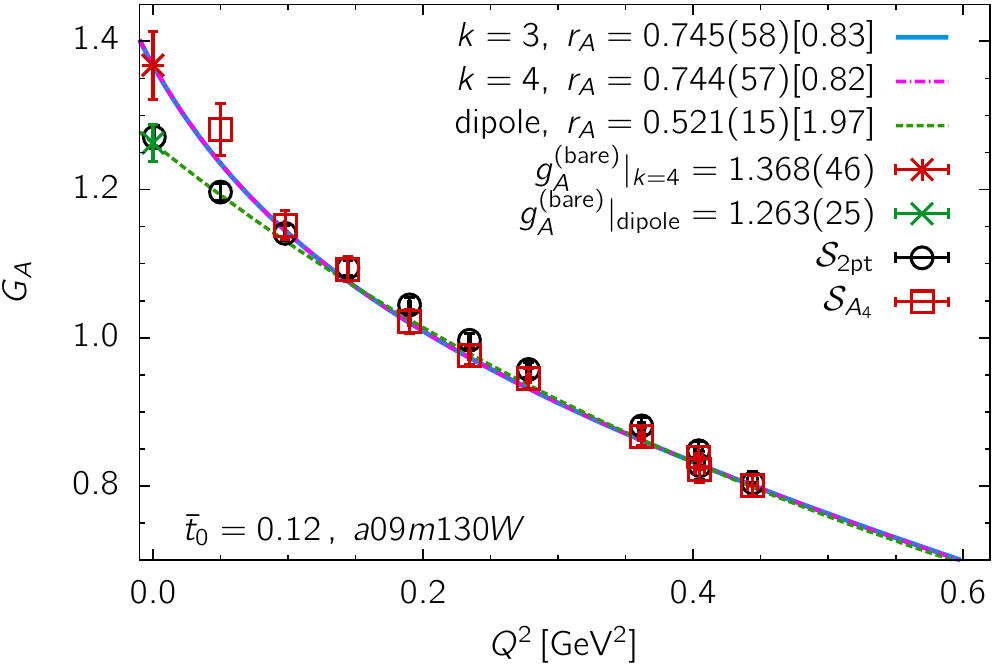}\hspace{2mm}
  \includegraphics[width=0.63\columnwidth]{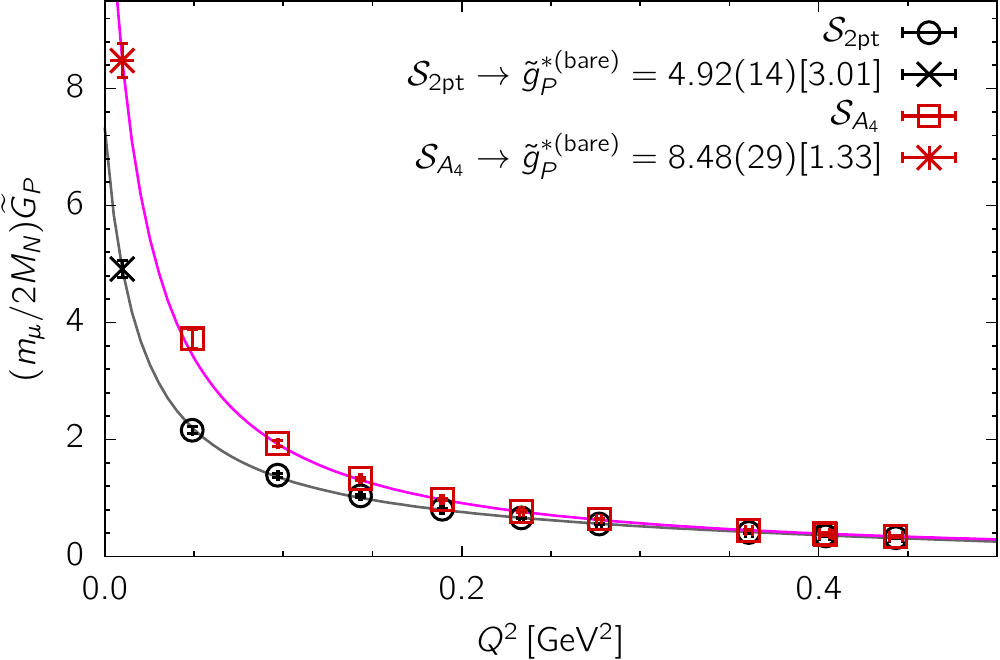}\hspace{2mm}
  \includegraphics[width=0.63\columnwidth]{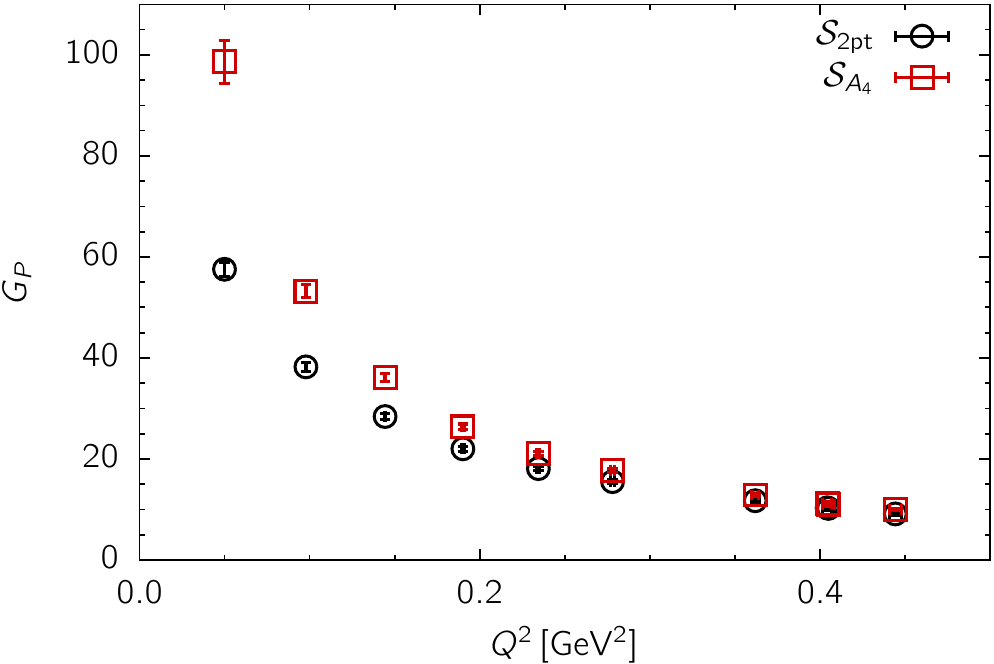}
  \caption{Comparison of $G_A$ (left),
    $(\widetilde{G}_P)$ (middle) and $G_P$ (right) versus $Q^2$, in
    units of GeV${}^2$, obtained using the two stategies ${\cal
      S}_{\rm A4}$ (red) and ${\cal S}_{\rm 2pt}$ (black). The lines show the dipole and $k^3$ and $k^4$ 
     $z$-expansion fits to $G_A$ and PPD ansatz to $\widetilde{G}_P$. 
}
  \label{fig:3FF}
\end{figure*}

The results for the three form factors $G_A$, $\widetilde{G}_P$ and
$G_P$ are compared in Fig.~\ref{fig:3FF}. The effect of using ${\cal
  S}_{\rm A4}$ is clear and largest for ${\bm n} = (1,0,0)$. Also, the
change in $G_A(Q^2)$ is only apparent for ${\bm n} = (1,0,0)$,
consequently data at smaller $Q^2$ are needed to quantify its $Q^2 \to
0$ limit.

The pattern, that the
effect increases as $Q^2 \to 0$, $a \to 0$ and $M_\pi \to M_\pi^{\rm physical}$, is
confirmed by the analysis of the 11 ensembles described in
Ref.~\cite{Jang:2019jkn}, and these detailed results will be presented
in a separate longer paper~\cite{PNDME:AFF2019}.

With $G_A$, $\widetilde{G}_P$ and $G_P$ in hand, we present the test
of the PCAC relation, Eq.~\eqref{eq:PCAC}, in Fig.~\ref{fig:PCAC}. The
figure also shows data for the pion-pole dominance (PPD) hypothesis
that relates $\widetilde{G}_P$ to $G_A$ as ${\tilde G}_P(Q^2) =
{G}_A(Q^2) \frac{4M^2}{Q^2+M_{\pi}^2}$. It is clear that both
relations are satisfied to within 5\% at all $Q^2$ with ${\cal S}_{\rm
  A4}$, whereas the deviation grows to about 40\% with ${\cal S}_{\rm
  2pt}$ at ${\bm n} = (1,0,0)$ as first pointed out in
Ref.~\cite{Rajan:2017lxk}.  What is also remarkable is that the PPD
relation with the expected proportionality factor $4M^2$ provided by
the Goldberger-Treiman relation~\cite{PhysRev.111.354} tracks the
improvement in PCAC. In fact, the data for the two tests overlap at
all $Q^2$.

\begin{figure*}[!htb]  
  \centering
  \subfigure{
  \includegraphics[width=0.95\columnwidth]{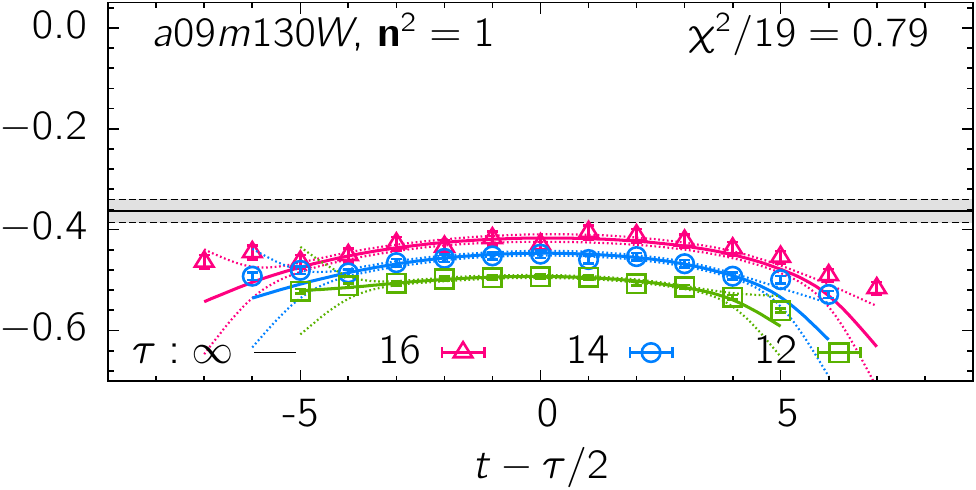}\hspace{6mm}
  \includegraphics[width=0.95\columnwidth]{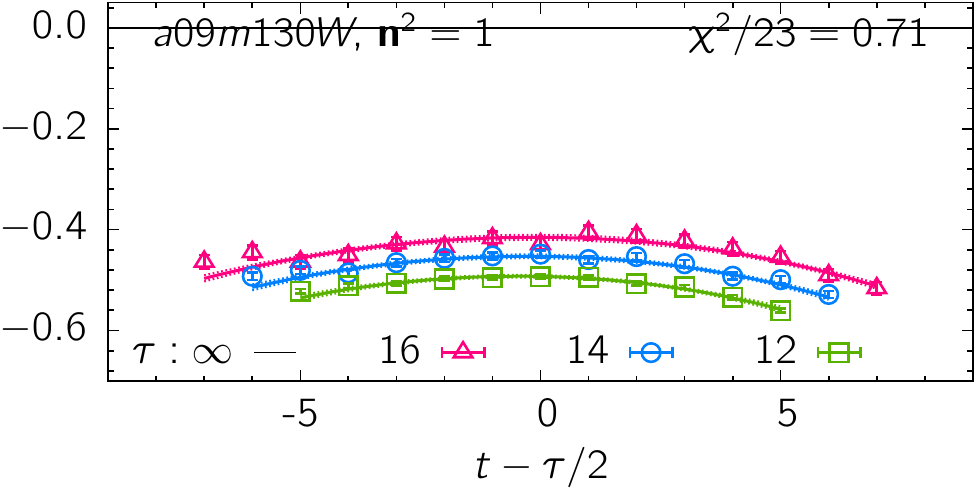}
  }
  \caption{Data and fits to $\partial_4 A_4 $ using ${\cal S}_{\rm
      2pt}$ (left) and ${\cal S}_{\rm A4}$ (right). The values of
    $\tau$ used and the $\chi^2/$DOF of the fits are given in the
    legend. The horizontal grey band showing the ground
    state value has tiny uncertainty in the right panel.  }
  \label{fig:d4A4}
\end{figure*}

The last test we perform is the relation $\partial_4 A_4 = (M-E)
A_4$ that should be satisfied by the ground state matrix element.  
The data and fits for $\partial_4 A_4$ are shown in Fig.~\ref{fig:d4A4}.  
The values of $(M-E) A_4$ are essentially zero in both cases; for ${\cal
  S}_{\rm 2pt}$ because $M-E$ is small. Again, it is clear
that the relation is only satisfied for ${\cal S}_{\rm A4}$.

The bottom line is that the two relations, PCAC and $\partial_4 A_4 =
(M-E) A_4$, and the pion-pole dominance hypothesis are all satisfied
using ${\cal S}_{\rm A4}$ but not with ${\cal S}_{\rm 2pt}$.  The data
shown in Fig.~\ref{fig:massgaps} is consistent with the picture that  the
``new'' lower energy state is mainly due to the current $A_\mu(\bm q)$
injecting a pion with momentum $\bm q$.  There are two open questions:
(i) how do we extract $g_A$, {\it ie}, what is the analogous lowest excited
state at zero momentum since we cannot determine its parameters from
the $A_4$ correlator, and (ii) why it was not clear from the data
shown in Figs.~\ref{fig:ESC} and~\ref{fig:ESCcomp} that the mass gaps
used in ${\cal S}_{\rm 2pt}$ were too large. These points are
addressed below.

Results for $g_A$ have been obtained from the $A_3$ correlator at
zero-momentum in all previous calculations because it has the best
signal. The states with the lowest energy that are candidates for the
$\frac{1}{2}({\frac{1}{2}}^+)$ excited state at zero momentum in this
correlator are $N \pi \pi$ and $N({\bm p}) \pi({\bm -p})$. Both of
these are lighter than the radial excitations N(1440) and N(1710) and
dominate their decay. Their relativistic non-interacting
energies, in a box of size $L/a=64$ used for the $a09m130W$ ensemble,
are about 1230~MeV ($\Delta M_1 a \approx 0.12$).  Our previous
argument favors $N \pi \pi$: the current $A_3 ({\bm p}=0)$ is more
likely to insert a $\pi \pi$ state at zero momentum, whereas in the
other case it would need to insert a pion with $({\bm p}=(1,0,0))$ and at the
same time cause the transition $N(0) \to N({\bm p}=(-1,0,0))$ to ensure
zero total momentum. In any case, since the only quantity that enters
in our analysis is the mass gap and not the specifics of the excited
state, we take the common value, $\Delta M = 1230$~MeV, in the
reanalysis of $A_3$ to extract $g_A$.

\begin{table}[!tbh]
  \caption{Final results from both strategies, ${\cal S}_{\rm A4}$ and
    ${\cal S}_{\rm 2pt}$. $g_A$ is obtained in three ways as discussed
    in the text, and $r_A$ and $g_P^\ast$ from $z^4$ fits. Note these 
    estimates are at fixed $a \approx 0.0871$~fm.}
  \label{tab:FINAL}
  \begin{ruledtabular}
    \begin{tabular}{l|c|c|c|c|c}
                      & $g_A|_{3pt}$ & $g_A|_{z{\rm -exp}}$  & $g_A|_{\rm dip}$  & $r_A$(fm) & $g_P^\ast$   \\
\hline                                                                      
${\cal S}_{\rm A4}$   & 1.30(6)      & 1.30(7)               & 1.20(6)           & 0.74(6)   &  8.06(44)    \\
\hline
${\cal S}_{\rm 2pt}$  & 1.25(2)      & 1.19(5)               & 1.20(5)           & 0.45(7)   &  4.67(24)    \\ 
    \end{tabular}
  \end{ruledtabular}
\end{table}

In what follows, all results for the renormalized axial current are
presented using $Z_A=0.95(4)$ taken from
Ref.~\cite{Gupta:2018qil}. Fits to the zero-momentum $A_3$ correlator
with prior $\Delta M a = 0.12(4)$ give $g_A$ in the range 1.29--1.31
depending on the values of $\tau$ used in the fit compared to
$g_A=1.25(2)$ using ${\cal S}_{\rm 2pt}$ given in
Ref.~\cite{Gupta:2018qil} (column 2 in Table~\ref{tab:FINAL}).
However, fits with priors in the range $0.1 \lesssim \Delta M_1
\lesssim 0.4$ are not distinguished on the basis of $\chi^2$/DOF.  The
output $\Delta M_1$ tracks the input prior, and the value of $g_A$
increases as the prior value is decreased.  Thus, we regard this
method as giving $g_A$ with uncontrolled systematics--the relevant
$\Delta M_1$ has to be determined first.  Parenthetically, similar
fits to extract the scalar and tensor charges $g_S$ and $g_T$ are much
more stable, the value of the output $\Delta M_1$ is far less
senstitive to the prior and the results vary by $\lesssim 2\sigma$ as
will be shown in Ref.~\cite{PNDME:AFF2019}.

Our current best estimate for $\Delta M_1$ on a given ensemble is to
take the lower of the $N \pi \pi$ or $N({\bm n}^1=1) \pi({\bm n}^2=1)$
states.  Assuming they are roughly degenerate, one can use the value
of $\Delta M_1^{A4}a \approx 0.1$ shown in
Fig.~\ref{fig:massgaps} at ${\bm n}^2= 1$ that, as we have argued above, corresponds to
the latter state. Using this $\Delta M_1^{A4}a$, our analysis of the
$A_3$ data gives $g_A = 1.30(6)$.

The second way we extract $g_A$ is to parameterize the $Q^2$
dependence of $G_A(Q^2)$ using the $z$-expansion and the dipole
ansatz. The $z$-expansion fits using the process defined in
Ref.~\cite{Rajan:2017lxk,Jang:2019jkn} give $g_A=1.30(7)$ for ${\cal
  S}_{\rm A4}$ compared to $g_A=1.19(5)$ using ${\cal S}_{\rm
  2pt}$. These results are independent of $k$ for $k > 2$ in the
$z^k$-expansion. The dipole fit gives $g_A=1.20(6)$ with a large
$\chi^2$/DOF = 1.97 and the results are essentially the same for
${\cal S}_{\rm A4}$ and ${\cal S}_{\rm 2pt}$ as can be seen in
Fig~\ref{fig:3FF}. One can fix the dipole fit to not miss the crucial
low $Q^2$ points by putting a cut on $Q^2$, however, for this study we
choose to neglect it.

The root-mean-squared charge radius extracted using the $z$-expansion
fits gives $r_A=0.74(6)$~fm with ${\cal S}_{\rm A4}$ and
$r_A=0.45(7)$~fm with ${\cal S}_{\rm 2pt}$.  Once the lattice data
have been extrapolated to the continuum limit, they can be compared
with (i) a weighted world average of (quasi)elastic neutrino and
antineutrino scattering data~\cite{Bernard:2001rs}, (ii) charged pion
electroproduction experiments~\cite{Bernard:2001rs}, and (iii) a
reanalysis of the deuterium target data~\cite{Meyer:2016oeg}:
\begin{eqnarray}
r_A&=& 0.666(17)~{\rm fm}   \qquad \nu,{\overline \nu}-{\rm scattering}    \,,  \nonumber \\
r_A&=& 0.639(10)~{\rm fm}   \qquad {\rm Electroproduction} \,,  \nonumber \\
r_A&=& 0.68(16)~{\rm fm}  \, \  \qquad {\rm Deuterium} \,,
\label{eq:rA_expt}
\end{eqnarray}

The induced pseudoscalar charge $g_P^\ast$, defined as 
\begin{equation} 
g_P^\ast \equiv  \frac{m_\mu}{2M} {\tilde G}_P(Q^2 = 0.88m_\mu^2)   \,,
\label{eq:GPstar}
\end{equation}
is obtained by fitting ${\tilde G}_P(Q^2)$ 
using the small $Q^2$ expansion of the PPD ansatz: 
\begin{equation}
\frac{m_\mu}{2 M}  \frac{{\tilde G}_P(Q^2)}{g_A} = \frac{c_1}{M_\pi^2 + Q^2} + c_2 +c_3 Q^2 \,,
\label{eq:PPDfit}
\end{equation}
Our result using ${\cal S}_{\rm A4}$ is $g_P^\ast = 8.06(44)$, while
the MuCap experiment gave $g_P^\ast|_{\rm MuCap} =
8.06(55)$~\cite{Andreev:2012fj,Andreev:2015evt}.

We caution the reader that all the results summarized in
Table~\ref{tab:FINAL} are at fixed $a \approx 0.0871$~fm.  Comparison to the
phenomenological values should be made only after extrapolation to the
continuum limit. The goal of this paper is to highlight the changes on
using ${\cal S}_{\rm A4}$.

\begin{figure}[tbp]
\centering
\includegraphics[width=0.97\linewidth]{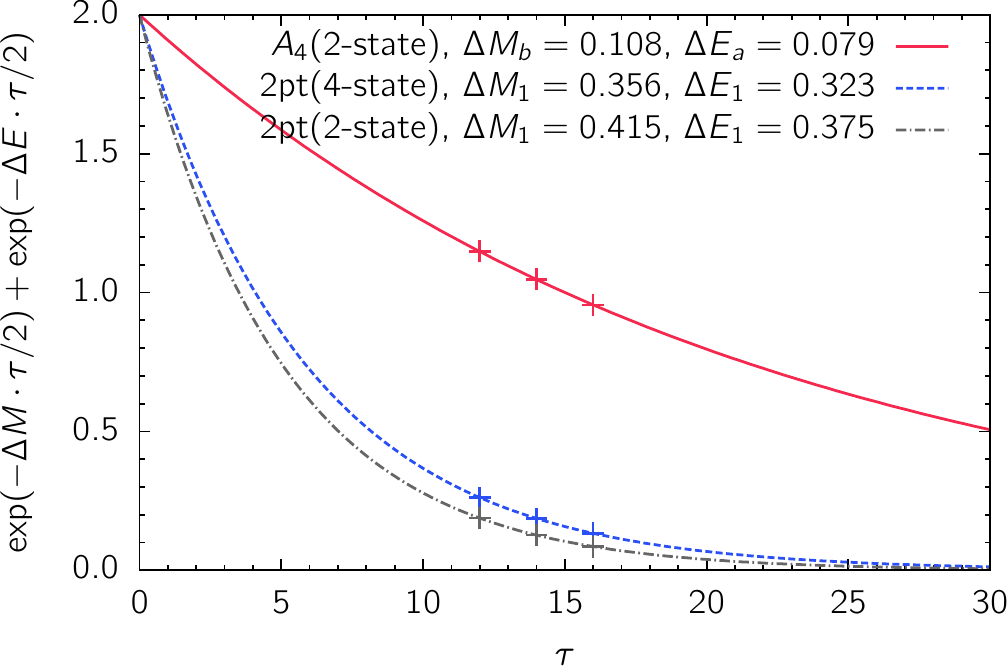}
\caption{Cartoon of the behavior of the ESC in 3-point functions evaluated at the midpoint $t=\tau/2$ 
for three typical values of $(M_1-M_0)$ and $(E_1^{{\bm p}} - E_0^{{\bm p}})$ 
as a function of the source-sink separation $\tau$. The plus symbols show the three values of $\tau/a=12, 14, 16$
at which the lattice data are presented. }
\label{fig:exp}
\end{figure}

Second, we comment on why the lower-energy state is missed when
following ${\cal S}_{\rm 2pt}$. It is well known that extracting $E_i$
from n-state fits to ${ C}^{\rm 2pt}$ gives $E_i$ with ESC since the
number of pre-plateau data point that are sensitive to excited states
are typically 8--12 as shown in Fig.~\ref{fig:2pt}. While we find an
$\approx 15\%$ change between a 2- and 4-state fit, we did not
anticipate $E_1^{\rm A4} \sim E_1^{\rm 2pt} /4$ at small $Q^2$ as
shown in Fig.~\ref{fig:exp}. The known methodology to getting a more
realistic excited state spectrum in a finite box with nucleon quantum
numbers is to construct a large basis of interpolating operators,
including operators overlapping primarily with multiparticle states,
and solve the generalized eigenvalue problem (GEVP)~\cite{Fox:1981xz}
in a variational
approach~\cite{Edwards:2011jj,Alexandrou:2013fsu,Lang:2013uwa,Dudek:2018wai,Briceno:2017max}.
One should then compare the energies with lattice data, for example in the axial case with $E_1^{\rm
  A4}$, to determine which states contribute to a given 3-point
function. This option will be explored in future
calculations.\looseness-1

To get a rough picture of the impact of the choice of $E_1$ on the ESC
in 3-point functions, assume that the prefactors in the two $0
\leftrightarrow 1$ transition terms are equal and unity, and the $1
\leftrightarrow 1$ term can be neglected in Eq.~\ref{eq:3pt}. Then the
ESC should fall off exponentially as $e^{-(M_1-M_0)(\tau/2)} +
e^{-(E_1 - E_0)(\tau/2)}$.  In Fig.~\ref{fig:exp}, we plot this
function for three typical values of $(M_1-M_0)$ and $(E_1 - E_0)$
with ${\bm n} = (0,0,1)$. These values are from the 2- and 4-state
fits to the 2-point function and those extracted from a 2-state fit to
the $A_4$ 3-point function.  Over the interval $10 < \tau/a < 16$,
corresponding to 0.9--1.4~fm in which lattice data are typically
collected, the exponential fall-off is approximately
linear. Furthermore, the three curves in this range can be roughly
aligned by a constant shift in their magnitude, {\it ie}, by just a change
in the prefactors we have set to unity in Eq.~\ref{eq:3pt}, and which
are free parameters in the actual fits.  Thus, over a limited range of
$\tau$, the expected exponential convergence can be masked to look
linear. On the other hand, the size of the ESC is very sensitive to
$E_1$ and large even at $\tau/a = 25$ for $E_1^{A_4}$.  The lesson is
that while the excited state energy gaps impact the magnitude of the
ESC at any given $\tau$, checks on the $E_i$ using 2-state fits and
the convergence to $\tau \to \infty$ of ground state matrix elements
is hard to judge from a limited range of $\tau$ even for very
different energy gaps. As discussed above, the extraction of $g_A$ is
plagued by this problem since we are not able to extract $M_1^{\rm A4}$ from a 3-point function. In
short, it is very important to determine $E_1$ reliably. Once this is
done, even 2-state fits give reasonable estimates of the $\tau \to
\infty$ value based on the consistency checks discussed above.

An attempt to resolve the PCAC conundrum has been presented
in~\cite{Bali:2018qus}.  We contend that it missed resolving the lower
energy state and did not solve the problem. The projected currents
$A_\mu^\perp$ and $P^\perp$ introduced in their work consist of a
rotation in the basis of the five currents $A_\mu$ and $P$.  For the
lattice ensembles and parameters explored in their~\cite{Bali:2018qus}
and our~\cite{Rajan:2017lxk,Jang:2019jkn} calculations, the three
$A_i^\perp$ essentially remain within the space of the $A_i$. Thus,
the ${\cal S}_{\rm 2pt}$ strategy with $A_i^\perp$ gives $G_A(Q^2)$
and $\widetilde{G}_P$ that are essentially unchanged, and one
continues to get a low value for $g_P^\ast$~\cite{Bali:2018qus}. The
operator $A_4^\perp$ is mostly rotated into the $A_i$.  Thus
$A_4^\perp$ no longer shows the large ESC, and the ``sinh'' behavior
illustrated in Fig.~\ref{fig:ESC} becomes ``cosh'' like. Their ``fix''
to PCAC comes from $P^\perp$, which now gets its dominant contribution
from $A_4$ and $\partial_4 A_4$. Analysis of our $a09m130W$ ensemble
shows that the contribution of the $\partial_4 A_4$ part is roughly
three times that of $P$ due to the small value of the PCAC mass
$\widehat m$ in the definition of $P^\perp$. Also, note that, by construction, the total contribution of
$(P^\perp-P)$ is supposed to be zero in the ground state.  On the other
hand, we contend that the solution to the PCAC problem lies in the
identification of the lower energy excited state[s] that, as we have
presented, should be used to remove the ESC in all 3-point
axial/pseudoscalar correlators. Using ${\cal S}_{\rm A4}$ changes the
results for all three form factors, especially at low $Q^2$.

\section{Conclusions}

All previous lattice calculations of the three form factors $G_A$,
$\widetilde{G}_P$ and $G_P$~\cite{Bali:2014nma,Rajan:2017lxk,Green:2017keo,Alexandrou:2017hac,Capitani:2017qpc,Ishikawa:2018rew,Shintani:2018ozy}, showed significant violations of the PCAC
relation, Eq.~\eqref{eq:PCAC}. This failure had cast doubts on the
lattice methodology for extracting these form factors. In this work,
we show that the systematic responsible for the violation is a lower
energy excited state missed in previous analyses. Furthermore, its
energy can be extracted from fits to the $A_4$ 3-point
function. Detailed analysis of the $A_4$ correlator had, so far, been
neglected as it is dominated by ESC and is not needed to extract the
form factors. Using the mass/energy gaps of this lower excited-state,
we show that lattice data satisfy PCAC to within 5\%, the level
expected with reasonable estimates of the current level of statistical
and systematic errors.  An additional consistency check is that the
ground state matrix elements now satisfy the relation $\partial A_4 =
(M-E) A_4$.  We also show that pion-pole dominance works to the same
level as PCAC with the proportionality constant $4M^2$ suggested by
the Goldberger-Treiman relation.

We show that the direct extraction of $g_A$ from the $A_3$ correlator
at zero-momentum requires knowing the energies of the excited states
that give the dominant contamination, {\it ie}, the result for $g_A$ is
particularly sensitive to the input value of the mass gap $\Delta
M_1$. We show that the $\Delta M_1$ obtained from the 2-point function
is much larger than what is expected, so alternate methods for
determining it are needed because fits to the $A_3$ correlator data,
while precise, are not able to distinguish between $\Delta M_1$ 
in a wide range. Our new analysis using two plausible estimates of 
$\Delta M_1$ gives $g_A=1.30(6)$. 

We provide heuristic reasons for why previous fits to remove ESC with
a large $\Delta M_1$ did not exhibit large $\chi^2/{\rm DOF}$, and why
the smaller values of the mass gaps that impact the extraction of the
form factors were missed.  For the form factors at $Q^2 \neq 0$, the
good news is that implementing this improvement, in the axial channels
(and an analogous procedure for the vector current for extracting
electromagnetic form factors), does not require the generation of new
lattice data but only a reanalysis.

We demonstrate the improvement in $G_A(Q^2)$, $\widetilde{G}_P(Q^2)$
and ${G}_P(Q^2)$ by analysing a physical mass ensemble with $a \approx
0.0871$~fm, $M_\pi \approx 138$~MeV~\cite{Rajan:2017lxk,Jang:2019jkn}.
We perform both the dipole and $z$-expansion fits to $G_A(Q^2)$ to
parameterize the $Q^2$ behavior and extract the axial charge radius
squared, $\rAsq$.  The dipole ansatz does not fit the data well and is
dropped.  The $z$-expansion fit gives $r_A = 0.74(6)$~fm.  We fit
$\widetilde{G}_P(Q^2)$ using the pion-pole dominance ansatz and find
$g_P^\ast = 8.06(44)$. To obtain results for these quantities in the
continuum limit, a full analysis of the 11 ensembles described in
Ref.~\cite{Jang:2019jkn} is in progress.

\section{Acknowledgement}
We thank the MILC Collaboration for providing the 2+1+1-flavor HISQ
lattices. The calculations used the Chroma software
suite~\cite{Edwards:2004sx}.  Simulations were carried out on computer
facilities of (i) the National Energy Research Scientific Computing
Center, a DOE Office of Science User Facility supported by the Office
of Science of the U.S. Department of Energy under Contract
No. DE-AC02-05CH11231; and, (ii) the Oak Ridge Leadership Computing
Facility at the Oak Ridge National Laboratory, which is supported by
the Office of Science of the U.S. Department of Energy under Contract
No. DE-AC05-00OR22725; (iii) the USQCD Collaboration, which are funded
by the Office of Science of the U.S. Department of Energy, and (iv)
Institutional Computing at Los Alamos National Laboratory.
T. Bhattacharya and R. Gupta were partly supported by the
U.S. Department of Energy, Office of Science, Office of High Energy
Physics under Contract No.~DE-AC52-06NA25396.  T. Bhattacharya,
R. Gupta, Y.-C. Jang and B.Yoon were partly supported by the LANL LDRD
program.  Y.-C. Jang is partly supported by the Exascale Computing
Project (17-SC-20-SC), a collaborative effort of the U.S. Department
of Energy Office of Science and the National Nuclear Security
Administration.


%
\bibliography{ref} 

\end{document}

%% file: macro.tex
\providecommand{\abs}[1]{\lvert#1\rvert}

\providecommand{\matrixe}[3]{\langle#1\lvert#2\rvert#3\rangle}

\providecommand{\MeV}{\mathrm{MeV}}

\providecommand{\fm}{\mathrm{fm}}

\providecommand{\CL}{\nonumber\\}

\newcommand{\Dslash}{\ensuremath{{D\kern -0.65em /}}}

\definecolor{green}{rgb}{0.1, 0.8, 0.1}